\documentclass[preprint,proceedings]{rmaa}

\suppressfulladdresses 

\usepackage{paralist}

\usepackage{psfrag,color}

\newcommand{\CS}[1]{\texttt{\textbackslash #1}}

\def\modot{\, {\rm M}_\odot}
  
\SetYear{2006}
\SetConfTitle{Circumstellar Media and Late Stages of Massive Stellar Evolution}

\title{Models for the circumstellar medium of long gamma-ray burst progenitor 
candidates} 

\author{
  A. J. van Marle,\altaffilmark{1} 
  N. Langer,\altaffilmark{2}
  A. Achterberg,\altaffilmark{2}
  and G. Garc{\a'i}a-Segura\altaffilmark{3}}

\altaffiltext{1}{Bartol Research Institute,
  University of Delaware, 102 Sharp Lab, 19716 DE,
  USA (A.vanMarle@astro.uu.nl).}

\altaffiltext{2}{Astronomical Institute, Utrecht University,
  P.O.Box 80\,000, 3508 NL-TA Utrecht, Netherlands (n.langer, 
A.Achterberg@astro.uu.nl).}

\altaffiltext{3}{Instituto de Astronom{\a'i}a-UNAM, APDO Postal 877, Ensenada, Baja 
California, Mexico (ggs@astrosen.unam.mx).}

\shortauthor{van Marle \etal}
\shorttitle{CSM of long GRB progenitor candidates}

\listofauthors{A. J. van Marle, N. Langer, A. Achterberg, G. Garc{\a'i}a-Segura}
\indexauthor{van Marle, A. J..}
\indexauthor{Langer, A.}
\indexauthor{Achterberg, A.}
\indexauthor{Garc{\a'i}a-Segura, G.}

\abstract{
 We present hydrodynamical models of circumstellar medium (CSM) of long gamma-ray burst 
(GRB) progenitor candidates. 
 These are massive stars that have lost a large amount of mass in the form of  stellar 
wind during their evolution. 

 There are two possible ways to probe the CSM of long GRB progenitors. \\
 Firstly, the GRB afterglow consists of synchrotron radiation, emitted when the GRB jet 
sweeps up the surrounding medium. 
 Therefore, the lightcurve is directly related to the density profile of the CSM.
 The density can either decrease with the radius squared (as is the case for a freely 
expanding stellar wind) or be constant (as we would expect for shocked wind or the 
interstellar medium). \\
 Secondly, material between the GRB and the observer will absorb part of the afterglow 
radiation, causing absorption lines in the afterglow spectrum. 
 In some cases, such absorption lines are blue-shifted relative to the source 
indicating that the material is moving away from the progenitor star. 
 This can be explained in terms of wind interactions in the CSM.
 We can use the CSM of these stars to investigate their prior evolutionary stage.

}

\addkeyword{gamma rays: bursts}
\addkeyword{Stars: winds, outflows}
\addkeyword{Stars: circumstellar matter}
\addkeyword{Stars: Wolf-Rayet}
\addkeyword{line: identification}
\addkeyword{hydrodynamics}
\addkeyword{methods: numerical}

\begin{document}
\maketitle

\section{Introduction}
\label{sec:intro}

\begin{figure}[!t]
\resizebox{\hsize}{!}{\includegraphics[width=\textwidth]{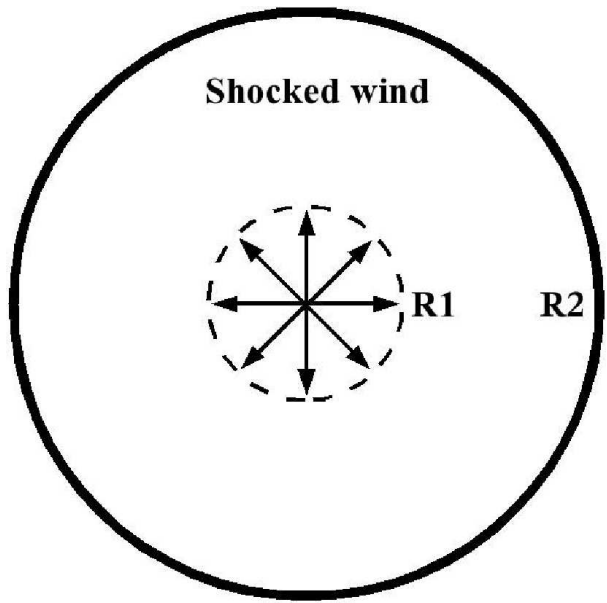}}
  \caption{Schematic view of a circumstellar bubble (not to scale).
       The free-streaming stellar wind  passes through the wind termination shock (R1) 
to enter the hot bubble of shocked
       wind material. The high thermal pressure in the hot bubble sweeps up a shell 
(R2), which expands into the ISM.}
  \label{fig:diagram}
\end{figure}

  The generally accepted model for long gamma-ray bursts (GRBs) is the collapsar model 
\citep{W93}. 
  According to this model a gamma-ray burst occurs if a rapidly rotating Wolf-Rayet 
star (the final evolutionary stage of a massive star)  collapses toward a black hole. 
  During the collapse, an accretion disk is formed around the black hole. 
  The GRB is produced by a relativistic jet that shoots out of the pole, driven by 
energy generated by the accretion onto the black hole. 
  As the jet moves outward it first encounters the outer envelope of the progenitor 
star. 
  Beyond this it moves into the circumstellar medium that has been formed by the 
stellar wind of the progenitor (see Fig.~\ref{fig:diagram}). The jet hits a region of 
free streaming stellar wind, then moves through the wind termination shock into the 
shocked wind region and eventually - provided it has sufficient energy to penetrate 
that far - into the interstellar medium (ISM) \citep{CLF04, EGDM05, RGSP05, MLAG06a, 
MLAG06b}.
  
 The GRB jet sweeps up the surrounding medium as it expands, accelerating particles to 
relativistic speeds. 
  As a result, synchrotron radiation is emitted, which we observe as the GRB afterglow. 
  There are two methods to use the afterglow radiation to investigate the surrounding 
medium.

\subsection{Circumstellar density profiles}
The afterglow lightcurve is directly related to the density profile of the surrounding 
medium. 
  Numerical models show that the density can either decrease with the radius squared or 
be constant \citep{CL00, PK01, PK02, CLF04}. 
  The latter is surprising for we would expect to find a free streaming wind region 
close to the star, which would show as $\rho~\propto~1/R^2$. 
  The most likely explanation for te occurance of a constant density medium is that the 
jet has already passed beyond the free-streaming wind region and has entered the 
shocked wind region where the density is indeed nearly constant \citep{W01,CLF04, 
MLAG06a, MLAG06b}. 
  However, this requires the wind termination shock to be extremely close to the star, 
which is difficult to understand in the case of a Wolf-Rayet stars because they tend to 
have powerful winds. 
  This i9s discussed in more detail in \CS{S}~\ref{sec:profile}.
\subsection{GRB afterglow absorption spectra}
   Circumstellar matter between the GRB and the observer will cause absorption lines in 
the GRB afterglow spectrum. 
  These lines provide information about the composition of the circumstellar medium 
(CSM).
  Some GRBs show the same line several times. 
  These lines are blue-shifted relative to the progenitor indicating a system of 
discrete velocities in the CSM. 
  This can be explained by hydrodynamical interactions between the stellar wind and the 
ISM or between different phases of the stellar wind \citep{MLG05a, MLG05b, MLAG06b}.
  We will discuss this in more detail in \CS{S}~\ref{sec:lines}.

\section{Forming a constant density medium close to a GRB}
\label{sec:profile}
  Most Wolf-Rayet stars have powerful stellar winds, which means that the wind 
termination shock lies far ($\sim~10~{\rm pc}$) from the progenitor star. 
  If this is the case, the GRB afterglow can never be generated in the shocked wind 
material, since the GRB jet will have ceased to be relativistic long before it has 
reached this distance. 
  If the afterglow is to be (partially) generated in the shocked wind, the termination 
shock has to be at $\lesssim~0.1~{\rm pc}$ \citep{CLF04}. 

\subsection{Outside influence}
    There are several possible causes for the wind termination shock to be close to the 
progenitor star, as was discussed in detail in our earlier papers \citep{MLAG06a, 
MLAG06b}. 
  Most of these scenarios involve an increase of the 'confining pressure', the pressure 
that restricts the expansion of the circumstellar bubble.
  For example, by increasing either the density or the thermal pressure of the ISM, the 
expansion rate of the circumstellar bubble decreases.
  Similarly, if the star has a supersonic velocity relative to the surrounding medium, 
the ram pressure of the stellar motion performs a similar function.
  Unfortunately, the effect of the confining pressure is limited.
  One needs an increase of three orders of magnitude to bring the wind termination 
shock one order of magnitude closer to the star \citep{MLAG06a}.
  This means either a very high density ($\sim~10^4~{\rm cm}^{-3}$) or a very high 
temperature ($\sim~10^8~{\rm K}$) in the ISM. 
  Stellar motion is more effective, especially since it can be combined with high ISM 
density. 
  Even so, the star has to move quite rapidly to create a bow shock (keeping in mind 
that it may be moving through an \ion{H}{II} region or a circumstellar bubble where the 
local sound speed is high because of the temperature).

\subsection{Internal influence}  
  Apart from these outside influences, it is also possible to move the wind termination 
shock closer to the star by decreasing the ram pressure of the stellar wind. 
  Since weak stellar winds usually occur at low metallicities \citep{NL00, EV06}, this 
would mean that GRBs showing a constant density in their afterglow lightcurve should 
occur more often at higher redshift.
    Finally, the penetration of the GRB jet into the surrounding medium depends 
directly on the energy of he GRB. Therefore, a powerful GRB can penetrate much deeper 
into the CSM and bring the wind termination shock within reach even at larger 
distances.

  \begin{figure}[!t]
  \centering
  \resizebox{\hsize}{!}{\includegraphics[width=\textwidth]{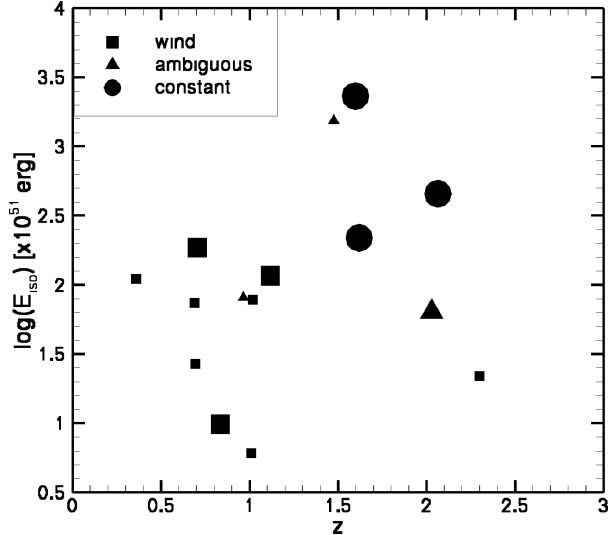}}
  \caption{Distribution of GRBs over the redshift-isotropic energy space published 
previously by \citep{MLAG06a}. GRBs whose afterglows show a constant density profile 
(circle) tend to have high energy and occur at larger redshifts than the afterglows 
which are formed in a free-streaming wind (square). 
The ones marked as triangles are ambiguous, which means that either different groups 
disagree on the nature of its afterglow, or no group has decided on its nature.
If the symbol is large, several groups have reached the same conclusion.
References for these afterglow models are: \citet{CL00}, \citet{PK01},    \citet{PK02} 
and \citet{CLF04}.}
  \label{fig:grbstat}
  \end{figure}

\subsection{Observational test}
  Density slopes around several GRBs have been determined from the afterglow light 
curves of a number of bursts \citep{CL00, PK01, PK02, CLF04}. Their distribution over 
the redshift-isotropic energy space (see Fig.~\ref{fig:grbstat}) seems to support our 
idea \citep{MLAG06a}. The GRBs with a constant density profile occur mostly at high 
redshift and have a high isotropic energy. (We use the isotropic energy of the burst, 
since it is irrelevant whether the burst has a high absolute energy, or a narrow beam. 
Both cause it to penetrate deeply into the surrounding medium.)
  
  If the afterglow is generated partially in the free streaming wind and partially in 
the shocked wind, as is quite feasible, one would expect to be able to observe the 
transition form one medium to another because of the jump in the local density. 
  However, recent work \citep{NG06} shows that the transition may not be visible.

\subsection{Alternative explanation}
  Another way to constrain the wind bubble is to suppose a period of extremely high 
mass loss before the onset of the Wolf-Rayet phase. 
  Such an outburst can be observed in the case of $\eta$ Carinae, which lost a 
considerable amount of mass ($\dot{M}~\gtrsim 10^{-2}~\modot/{\rm yr}$) during the 
nineteenth century \citep{D87}.
  Whether such a shell would actually constrain the Wolf-Rayet wind bubble is not 
certain. 
  Hydrodynamical calculations \citep{GML96, MLG05b} show that a shell, driven by a 
Wolf-Rayet wind, can break quite easily through the shell of shocked Red Supergiant 
wind material sitting at the wind termination shock. 
  The latter shell contains 10-20 $\modot$ but apparently does not impede the 
Wolf-Rayet wind driven expansion.

\section{Blue-shifted absorption lines in GRB afterglow spectra}
\label{sec:lines}
\begin{figure}[!t]
   \centering
  \resizebox{\hsize}{!}{\includegraphics[angle=-90,width=\textwidth]{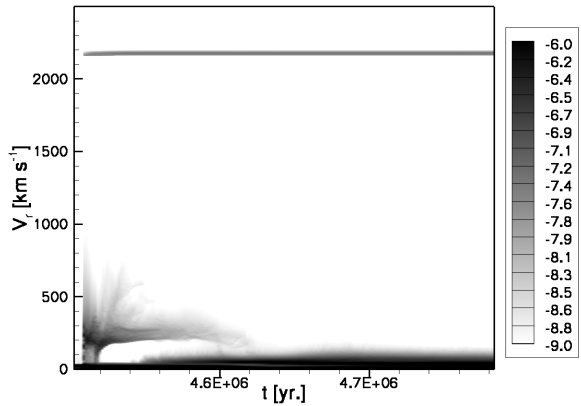}}      
  \caption{The column density in [${\rm g}/{\rm cm}^2$] as a function of radial 
velocity for a 40 M$_{\odot}$ star during the Wolf-Rayet stage \citep{MLG05a, MLG05b}. 
The zero velocity component is always present, as the interstellar medium itself 
provides gas that is standing still relative to the star.
    The Wolf-Rayet wind component at $2200~{\rm km/s}$ is visible during the entire 
Wolf-Rayet period. 
    However, the $150...650~{\rm km/s}$ component, caused by the Wolf-Rayet wind driven 
shell is short-lived.
    The column density in this figure is the average result for 200 radial gridlines.}
         \label{fig:coldens}
   \end{figure}

  Some GRBs show a series of absorption lines that are blue-shifted relative to the 
progenitor; notably GRB~021004, which shows at least {\bf six} absorption lines both in 
\ion{C}{IV} and \ion{Si}{IV} \citep{Fietal05, Setal05}. 
  These absorption lines show that there is a complicated system of discreet velocity 
features between the GRB and the observer. 
  In \citet{MLG05a, MLG05b}, we showed that these absorption features can be explained 
by stellar wind interactions that take place in the CSM of the progenitor star, using a 
40~M$_\odot$ star as described by \citet{SSMM92}. 
  The fastest absorption features ($\sim~3000~{\rm km/s}$) correspond to the wind from 
the Wolf-Rayet star, whereas the slower features ($\sim~150-650~{\rm km/s}$) are caused 
by fragments of the shell that the Wolf-Rayet wind swept up into the surrounding 
medium.
  The presence of these intermediate velocity components places a time constraint on 
the evolution of the progenitor star.
  They are comparatively short lived ($\lesssim~5\times10^4~{\rm yrs}$), since they 
will eventually dissipate into the surrounding gas (see Fig.~\ref{fig:coldens}).
  This means that the progenitor of GRB~021004 must have had a short Wolf-Rayet phase 
prior to its explosion. 

\subsection{Photo-ionization}
    The comparatively low ionization states of the ions that produce the absorption 
lines is not so easy to explain. 
  If the ions are in the direct line of the GRB, one would expect the atoms to be at 
high ionization states out to very large radii \citep{PCB06, Letal06}. 
  However, evidence so far suggests that GRB~021004 had a very strong wind, with a high 
density \citep{Letal06}. 
  This would put the wind termination shock far away from the star. 
  Alternatively, the wind might be highly clumped with optically thick clumps shielding 
part of the material from the ionizing photons \citep{Cetal06}.
  A second explanation lies in the possibility that the GRB jet is not homogeneous. 
  This may mean that the $\gamma$-radiation of the burst passed through a smaller area 
than the afterglow. 
  Therefore, the material that is absorbing the afterglow was not photo-ionized by the 
burst \citep{Setal05, Hetal06}. Even so, the wind termination shock could not have been 
very close to the progenitor star, since the afterglow itself produces enough high 
energy photons to ionize the gas to high ionization states at shorter range  
\citep{PCB06}.

\subsection{Alternative explanations} 
    It was suggested by \citet{Metal03} that radiative acceleration could account for 
the presence of blue-shifted absorption lines in a GRB afterglow spectrum.
  However, for the radiative force on the particles to be that powerful one would 
expect higher ionization states than those observed for GRB~021004 \citep{Setal05}. 
  Nor can this explanation account for the large number of discreet velocity features.
  Alternatively, the absorption features could be interstellar rather than 
circumstellar in origin. 
  E.g. the intermediate velocity lines could be caused by an expanding superbubble. 
  The velocities are rather high, for superbubbles in local galaxies typically have 
expansion velocities of less than $150~{\rm km/s}$ \citep{H97, TB88, M98}, but 
certainly not impossible if the progenitor is sitting inside a starburst region.
  However, the chances of a single GRB progenitor sitting at the center of three 
expanding superbubbles, each with a different velocity are not very large. 
  While a superbubble shell may account for one of the absorption features it is 
unlikely to account for all of them.
  Similarly, the high velocity components of the absorption spectrum can be a galactic 
wind rather than the stellar wind \citep{PCB06}. 
  While this would eliminate the problem of the low ionization states mentioned before, 
a galactic wind should produce an asymmetric absorption feature since the acceleration 
phase of the wind would absorb part of the radiation. 
  Moreover, it is difficult to explain two discreet absorption components at high 
velocities in this fashion.

\subsection{A possible supernova connection}
    Recently, three GRBs were detected that showed absorption lines at very high 
blue-shift relative to the progenitor: GRBs 050730, 050922 and 060418 \citep{Detal06, 
Petal06a, Petal06b}. 
  If the lines observed in these afterglows are indeed produced in the GRB progenitor 
host galaxy, the relative velocity of the gas would be of the order of 
$10^4...10^5~{\rm km/s}$.
  Such velocities would be very difficult to explain in terms of Wolf-Rayet wind 
velocities, which are usually at least an order of magnitude lower. 
  This may indicate, that these lines are just foreground noise, occurring in a 
different galaxy with a lower redshift than the GRB host galaxy. 
  However, there is another possible explanation. 
  Velocities such as these do occasionally occur in supernova explosions, as observed 
in blue-shifted absorption lines in the spectra of Type II supernova SN2005cs 
\citep{Pasetal06} and Type Ib supernova SN1987M \citep{Eetal04}.
  See also \citet{MK99} and \citet{CF06}.
  For the supernova ejecta to appear in absorption in the GRB afterglow, the supernova 
has to happen well in advance of the GRB. 
  This is not part of the normal collapsar model, but the possibility was suggested by 
\citet{VS98}, who proposed a 'supranova' model, where a supermassive neutron star falls 
back into a black hole after the original supernova. 
  The time interval between the two events could be as large as several years. 
  If some of the high velocity absorption lines observed in GRBs 050730, 050922 and 
060418 are indeed caused by the 'supranova ejecta' this would be a considerable step 
forward in our understanding of GRB formation.
  
\section{Discussion}
  The circumstellar medium of a massive star carries the fingerprints of the previous 
stages of the stellar evolution.
  It provides an excellent opportunity to investigate the evolutionary past of the 
central star.
  We have presented two methods for analyzing the CSM of GRB progenitors. 
  Through blue-shifted absorption lines in the GRB afterglow spectra we can not only 
study the composition of the CSM, but also the hydrodynamical interactions that took 
place, which in turn give us clues about the wind parameters of the previous 
evolutionary phases of the progenitor star.
  The density profiles of the CSM provide us with information on the balance between 
the ram pressure of the stellar wind and the confining pressure of the ISM. 
  There is no reason, why these tools can not be used in other situations as well.
  E.g., on supernovae and Wolf-Rayet stars. 
  If the CSM of these objects can be matched to the CSM around GRBs, it will be a major 
step forward in our understanding of the evolution of GRB progenitor stars.
  In the future we hope to provide similar models of the circumstellar medium around 
rapidly rotating stars \citep{YL05, YLN06, WH06}, which are thought to be GRB 
progenitors.

\acknowledgements
  A.J.v.M. acknowledges helpful discussions with H.-W. Chen, J.J. Eldridge and J.X. 
Prochaska.
      This work was sponsored by the Stichting Nationale Computerfaciliteiten (National 
Computing Facilities Foundation, NCF), with financial support from the Nederlandse 
Organisatie voor Wetenschappelijk Onderzoek (Netherlands Organization for Scientific 
research, NWO).
      This research was done as part of the AstroHydro3D project:\\
      (http://www.strw.leidenuniv.nl/AstroHydro3D/)

\end{document}